\documentclass[fleqn,usenatbib]{mnras}

\usepackage{newtxtext,newtxmath}

\usepackage[T1]{fontenc}
\usepackage{subfigure}
\DeclareRobustCommand{\VAN}[3]{#2}
\let\VANthebibliography\thebibliography
\def\thebibliography{\DeclareRobustCommand{\VAN}[3]{##3}\VANthebibliography}

\usepackage{graphicx}	
\usepackage{amsmath}	
\usepackage[usenames]{color}
\usepackage{xspace}	
\usepackage{float}

\def\gtsima{$\; \buildrel > \over \sim \;$}
\def\ltsima{$\; \buildrel < \over \sim \;$}
\def\prosima{$\; \buildrel \propto \over \sim \;$}
\def\gsim{\lower.7ex\hbox{\gtsima}}
\def\lsim{\lower.7ex\hbox{\ltsima}}
\def\simgt{\lower.7ex\hbox{\gtsima}}
\def\simlt{\lower.7ex\hbox{\ltsima}}
\def\simpr{\lower.7ex\hbox{\prosima}}

\newcommand{\kpch}{{h^{-1}{\rm kpc}}}
\newcommand{\mpch}{h^{-1}{\rm {Mpc}}}

\newcommand{\myr}{{\rm Myr}}
\newcommand{\gyr}{{\rm Gyr}}

\newcommand{\msunh}{h^{-1} M_\odot}

\newcommand{\LCDM}{$\Lambda$CDM\xspace}

\newcommand{\rhom}{\rho_{\rm m}}

\def\sparta{\textsc{Sparta}\xspace}
\def\moria{\textsc{Moria}\xspace}

\def\rockstar{\textsc{Rockstar}\xspace}
\def\consistenttrees{\textsc{Consistent-Trees}\xspace}

\def\planck{Planck\xspace}
\def\wmap{WMAP7\xspace}

\def\erebos{Erebos\xspace}

\def\rmm{{\rm m}}

\def\rmp{{\rm p}}

\def\rs{r_{\rm s}}

\def\mvir{M_{\rm vir}}
\def\rvir{R_{\rm vir}}

\def\mtom{M_{\rm 200m}}
\def\rtom{R_{\rm 200m}}

\def\rsp{R_{\rm sp}}

\def\xsp{X_{\rm sp}}

\usepackage{etoolbox}
\makeatletter
\patchcmd{\NAT@citex}
  {\@citea\NAT@hyper@{\NAT@nmfmt{\NAT@nm}\NAT@date}}
  {\@citea\NAT@nmfmt{\NAT@nm}\NAT@hyper@{\NAT@date}}
  {}
  {}
\patchcmd{\NAT@citex}
  {\@citea\NAT@hyper@{%
     \NAT@nmfmt{\NAT@nm}%
     \hyper@natlinkbreak{\NAT@aysep\NAT@spacechar}{\@citeb\@extra@b@citeb}%
     \NAT@date}}
  {\@citea\NAT@nmfmt{\NAT@nm}%
   \NAT@aysep\NAT@spacechar%
   \NAT@hyper@{\NAT@date}}
  {}
  {}
\patchcmd{\NAT@citex}
  {\@citea\NAT@hyper@{%
     \NAT@nmfmt{\NAT@nm}%
     \hyper@natlinkbreak{\NAT@spacechar\NAT@@open\if*#1*\else#1\NAT@spacechar\fi}%
       {\@citeb\@extra@b@citeb}%
     \NAT@date}}
  {\@citea\NAT@nmfmt{\NAT@nm}%
   \NAT@spacechar\NAT@@open\if*#1*\else#1\NAT@spacechar\fi%
   \NAT@hyper@{\NAT@date}}
  {}
  {}
\makeatother

\definecolor{orange}{rgb}{0.9,0.25,0}
\definecolor{purple}{rgb}{0.57,0.36,0.51}

\title[What sets the splashback radius?]{What sets the splashback radius of dark matter haloes: accretion history or other properties?}

\author[Shin \& Diemer]{
Tae-hyeon Shin,$^{1}$\thanks{E-mail: tae-hyeon.shin@stonybrook.edu}
Benedikt Diemer$^{2}$
\\
$^{1}$Department of Physics and Astronomy, Stony Brook University, Stony Brook, NY 11794, USA\\
$^{2}$Department of Astronomy, University of Maryland, College Park, MD 20742, USA\\
}

\date{}

\pubyear{2022}

\begin{document}
\label{firstpage}
\pagerange{\pageref{firstpage}--\pageref{lastpage}}
\maketitle

\begin{abstract}
The density profiles of dark matter haloes contain rich information about their growth history and physical properties. One particularly interesting region is the splashback radius, $\rsp$, which marks the transition between particles orbiting in the halo and particles undergoing first infall. While the dependence of $\rsp$ on the recent accretion rate is well established and theoretically expected, it is not clear exactly what parts of the accretion history $\rsp$ responds to, and what other halo properties might additionally influence its position. We comprehensively investigate these questions by correlating the dynamically measured splashback radii of a large set of simulated haloes with their individual growth histories as well as their structural, dynamical, and environmental properties. We find that $\rsp$ is sensitive to the accretion over one crossing time but largely insensitive to the prior history (in contrast to concentration, which probes earlier epochs). All secondary correlations are much weaker, but we discern a relatively higher $\rsp$ in less massive, older, more elliptical, and more tidally deformed haloes. Despite these minor influences, we conclude that the splashback radius is a clean indicator of a halo's growth over the past dynamical time. We predict that the magnitude gap should be a promising observable indicator of a halo's accretion rate and splashback radius.
\end{abstract}



\section{Introduction}
\label{sec:intro}

In the hierarchical picture of structure formation, dark matter haloes grow by accreting a mixture of smaller haloes and smooth matter (depending on the nature of the dark matter particles; e.g., \citealt{lacey_93}). We observe indirect evidence for this process in the form of rising galaxy luminosity functions \citep[e.g.,][]{muzzin_13}, higher clustering amplitudes at low redshift \citep[e.g.,][]{reid_12}, and the average infall velocities around haloes \citep[redshift-space distortions, e.g.,][]{percival_09} --- but the actual process of accretion onto individual haloes remains invisible. 

However, dark matter haloes carry an inevitable dynamical signature of their accretion: a caustic that forms at the apocentre of particles on their first orbit after infall. In idealised, spherically symmetric shell models, the caustic manifests as an infinitely sharp drop in density whose position depends only on the mass accretion rate \citep{Fillmore84, bertschinger_85, shi_16_rsp}. The `splashback radius,' $\rsp$, has also been found in simulations \citep{DK14, Adhikari14, MDK15}, where it is smoothed out due to the complexities of halo structure, as well as in satellite density and weak lensing observations \citep{more_16, Baxter17, Chang18, Zuercher19, Shin19, Shin21, Murata20, Adhikari21}. These detections open the possibility of constraining halo properties (such as the accretion rate) from the position of the splashback feature, or from the density profiles in general \citep{xhakaj_20, xhakaj_22}.

Before we can attempt to infer accretion rates from such observations, we need to understand which of many possible halo properties besides the accretion rate might affect the splashback radius: the full mass accretion history (MAH) is more complicated than the power-law growth envisioned in spherical collapse models \citep{wechsler_02, tasitsiomi_04_clusterprof, mcbride_09}; dark energy provides an effective force that pulls particles away from the halo centre; non-sphericity and tidal forces from nearby neighbours might change particle orbits; and the orbits are not purely radial as envision in one-dimensional shell models. Given these complications, it is not clear yet how `clean' a probe of the accretion rate $\rsp$ really is, and what definition of the accretion rate it is most sensitive to. Past work has been restricted to the dependencies of $\rsp$ on halo mass, redshift, and cosmology via $\Omega_{\rm m}$ \citep{diemer_17_rsp, diemer_20_catalogs}, as well as a relatively narrow range of definitions of the accretion rate \citep{DK14, lau_15}.

In this paper, we investigate a much larger array of possible correlations, including the full mass accretion history, different definitions of the recent accretion rate, structural parameters such as concentration and ellipticity, and dynamical parameters such as angular momentum. We focus on the normalised splashback radius to scale out the absolute halo mass and size, $\xsp \equiv \rsp / \rtom$, where $\rtom$ encloses an average overdensity of $200\ \rhom(z)$. The splashback mass generally follows similar, but weaker, trends compared to $\rsp$ \citep{diemer_17_rsp}. For individual haloes, $\rsp$ cannot be measured from the noisy density profiles but must instead be inferred from the 3D density structure \citep{Mansfield17} or from particle dynamics \citep{diemer_17_sparta}. We opt for the latter because it has produced the most comprehensive splashback halo catalogues to date \citep{diemer_20_catalogs}, and because we are interested in how halo physics influences the underlying particle orbits. For comparison, the radius where the profile slope is steepest is commonly used in the observations discussed above, but suffers from a complex and stochastic connection to the underlying dynamics \citep{oneil_21, wang_22_rsp_elucid}. We limit the scope of our investigation to realistic \LCDM cosmologies, although $\rsp$ also exhibits interesting dependencies on the nature of dark matter and dark energy \citep{adhikari_18_mog, contigiani_19_symmetron, banerjee_20}. Furthermore, we do not concern ourselves with whether the halo properties in question are observable --- most of them are not, at least not directly. However, we do show that $\rsp$ primarily responds to the recent mass accretion rate, which means that the dependencies on other quantities can mostly be neglected.

The paper is organised as follows. In Section~\ref{sec:methods}, we describe the simulations and methods. We analyse the correlations between the splashback radius and mass accretion history in Sections~\ref{sec:corr_rsp_mar}, and those with other halo properties in Section~\ref{sec:partial}. We close with a conclusion in Section \ref{sec:conclusion}. In Appendix~\ref{sec:app:rspdefs} we show that the exact definition of $\rsp$ does not affect our conclusions. In Appendix~\ref{sec:app:pca} we present a principal component analysis of mass accretion histories.

\begin{table*}
\centering
\caption{Overview of the $N$--body simulations used in this paper, which form a subset of the \erebos suite. $L$ denotes the box size in comoving units, $N^3$ the number of particles, $m_{\rm p}$ the particle mass, and $\epsilon$ the force softening length in comoving units. The redshift range of each simulation is determined by the first and last redshifts $z_{\rm initial}$ and $z_{\rm final}$, but snapshots were output only between $z_{\rm f-snap}$ and $z_{\rm final}$. The earliest snapshots of some simulations do not yet contain any haloes, so that the first catalogue with haloes is output at $z_{\rm f-cat}$; the \sparta and \moria data also begin at that redshift. The cosmological parameters are given in Section~\ref{sec:methods:sim}. Other simulation details, such as our system for choosing force resolutions, are discussed in \citet{DK14, diemer_15}.}
\label{table:sims}
\begin{tabular}{lcccccccccccc}
\hline
Name & $L$ & $N^3$ & $m_{\rm p}$ & $\epsilon$ & $\epsilon$ &$z_{\rm initial}$ & $z_{\rm final}$ & $N_{\rm snaps}$ & $z_{\rm f-snap}$ & $z_{\rm f-cat}$ & Cosmology \\
 & $(\mpch)$ & & $(\msunh)$ & $(\kpch)$ & $(L / N)$ & & & & & & & \\
\hline
L2000-WMAP7 & $2000$ & $1024^3$ & $5.6 \times 10^{11}$  & $65$  & $1/30$ & $49$ & $0$ & $100$ & $20$ & $4.2$ & WMAP7 \\
L1000-WMAP7 & $1000$ & $1024^3$ & $7.0 \times 10^{10}$ & $33$ & $1/30$ & $49$ & $0$ &  $100$ & $20$ & $6.2$ & WMAP7 \\
L0500-WMAP7 & $500$  & $1024^3$ & $8.7 \times 10^{9}$  & $14$ & $1/35$  & $49$ & $0$ &  $100$ & $20$ & $8.8$ & WMAP7 \\
L0250-WMAP7 & $250$  & $1024^3$ & $1.1 \times 10^{9}$  & $5.8$  & $1/42$  & $49$ & $0$ &  $100$ & $20$ & $11.5$ & WMAP7 \\
L0125-WMAP7 & $125$  & $1024^3$ & $1.4 \times 10^{8}$  & $2.4$  & $1/51$  & $49$ & $0$ &  $100$ & $20$ & $14.5$ & WMAP7 \\
L0063-WMAP7 & $62.5$ & $1024^3$ & $1.7 \times 10^{7}$  & $1.0$  & $1/60$ & $49$ & $0$ &  $100$ & $20$ & $17.6$ & WMAP7 \\
L0500-Planck & $500$  & $1024^3$ & $1.0 \times 10^{10}$  & $14$ & $1/35$  & $49$ & $0$ &  $100$ & $20$ & $9.1$ & Planck \\
L0250-Planck & $250$  & $1024^3$ & $1.3 \times 10^{9}$  & $5.8$  & $1/42$  & $49$ & $0$ &  $100$ & $20$ & $12.3$ & Planck & \\
L0125-Planck & $125$  & $1024^3$ & $1.6 \times 10^{8}$  & $2.4$  & $1/51$  & $49$ & $0$ &  $100$ & $20$ & $15.5$ & Planck & \\
\hline
\end{tabular}
\end{table*}


\section{Simulations and methodology}
\label{sec:methods}

\subsection{N-body simulations and Halo finding}
\label{sec:methods:sim}

We use a subset of the \erebos suite of dissipationless $N$-body simulations (Table~\ref{table:sims}). This suite contains simulations of different box sizes of two \LCDM cosmologies; we focus on the first, which is the same as that of the Bolshoi simulation \citep{klypin_11} and is consistent with \wmap \citep{komatsu_11}, namely, flat \LCDM with $\Omega_{\rm m} = 0.27$, $\Omega_{\rm b} = 0.0469$, $\sigma_8 = 0.82$, and $n_{\rm s} = 0.95$. The second is a \planck-like cosmology with $\Omega_{\rm m} = 0.32$, $\Omega_{\rm b} = 0.0491$, $h = 0.67$, $\sigma_8 = 0.834$, and $n_{\rm s} = 0.9624$ \citep[][]{planck_14}. The power spectra that correspond to these values were generated using \textsc{Camb} \citep{lewis_00}, and they were translated into initial conditions using the \textsc{2LPTic} code \citep{crocce_06}. The simulations were run with \textsc{Gadget2} \citep{springel_05_gadget2}. 

We identify haloes and subhaloes using the phase--space friends-of-friends \citep{davis_85} halo finder \textsc{Rockstar} \citep{behroozi_13_rockstar}. The halo catalogues computed at different redshifts (Table~\ref{table:sims}) are the combined into time-connected merger trees using the \textsc{Consistent-Trees} code \citep{behroozi_13_trees}. The halo catalogues and trees are described in detail in \citet{diemer_20_catalogs}. 

\subsection{Determining the splashback radius with SPARTA}
\label{sec:methods:sparta}

While \rockstar computes most of the halo properties we investigate in this paper, our focus is the splashback radius. While this halo boundary is apparent as a sharp drop in the average density profiles of haloes, measuring it reliably for individual haloes demands more advanced algorithms that rely either on the full three-dimensional density field \citep{Mansfield17} or on particle dynamics. We opt for the latter and use the \sparta code, a flexible, parallel framework for the dynamical analysis of particle-based astrophysical simulations \citep{diemer_17_sparta, diemer_20_catalogs}. \sparta follows each virtual particle as it falls into a halo and records the time and location of its first apocentre. It then reconstructs the halo's splashback radius by smoothing the distribution of the particle apocentres in time and taking its mean (denoted $R_{\rm sp,mn}$) or higher percentiles (e.g., $R_{\rm sp,90\%}$ for the 90th percentile). Similarly, mass definitions such as $M_{\rm sp,mn}$ and $M_{\rm sp,90\%}$ are computed as percentiles of the mass enclosed within the individual particle splashback events. 

Besides changes in the radius and mass of host (isolated) haloes, using a different radius definition also changes which haloes we call subhaloes. The \moria extension to \sparta computes these new host-subhalo definitions and combines the results from \rockstar, \consistenttrees, and \sparta in a single, convenient format for halo catalogues and merger trees. All data used in this paper is publicly available; we refer the reader to \citet{diemer_17_sparta} and \citet{diemer_20_catalogs} for details on the \sparta and \moria codes.

The different measures of the apocentre distribution provide different definitions of the splashback radius. For simplicity, we focus on $R_{\rm sp,75\%}$ in this work, but Figure~\ref{fig:cov_mat} and Appendix~\ref{sec:app:rspdefs} demonstrate that the correlations with other definitions are qualitatively very similar.

\subsection{Halo selection}
\label{sec:methods:data_sel}

For each cosmology, we combine the boxes listed in Table~\ref{table:sims}, i.e., box sizes between $62.5$ and $2000\ \mpch$ for the fiducial \wmap cosmology and a somewhat smaller range for the \planck cosmology. We focus on the 45 snapshots between $z=4.0$ to $z=0.13$ that are present in all \wmap simulations. We do not compute results at $z = 0$ because the \sparta algorithm becomes less accurate for the final few snapshots of a simulation \citep{diemer_17_rsp}. We combine haloes from the different boxes into a single sample per cosmology and per redshift, irrespective of which simulation box they originated from. We have confirmed that the results from the \planck cosmology are entirely consistent with those based on the \wmap simulation set. We thus report only results from the latter throughout the paper.

We make a number of cuts on the combined catalogues. First, we select only parent haloes using $R_{\rm sp,75\%}$ as the halo boundary. Second, we demand that haloes contain at least 1000 particles within $R_{\rm 200m}$. Third, the \sparta algorithm must have successfully computed splashback radii; this criterion is not fulfilled for a small number of edge cases \citep{diemer_20_catalogs}. Fourth, we exclude haloes whose total mass $M_{\rm 200m}$ is larger than twice the gravitationally bound mass within the same radius. Most of these haloes are experiencing significant tidal disruption from nearby neighbours \citep{diemer_22_prof1}.

Finally, we remove so-called fly-by haloes from the sample: haloes that were previously a subhalo but that currently orbit outside of the halo boundary \citep[e.g.,][]{balogh_00, mamon_04, gill_05}. While the splashback radius should, in principle, contain all orbiting material, our chosen definition of $R_{\rm sp,75\%}$ means that we expect 25\% of particle apocentres to occur outside of this radius, and a similar number of subhalo apocentres as well. While the fraction of fly-by haloes is smaller for splashback definitions than for $\rtom$, it can still be significant \citep{flyby}. The mass accretion histories of fly-by haloes are typically strongly influenced by their time as a subhalo and would confuse our results. We thus exclude all haloes that were classified as a subhalo in the past and whose past host still exists as a host halo. 

\subsection{Definitions of halo properties}
\label{sec:methods:defvar}

We consider a wide range of halo properties, which we express as dimensionless quantities to remove any dependencies on units or on the overall size scale of the halo. First, we normalise the splashback radius by $R_{\rm 200m}$ and consider $X_{\rm Rsp,75\%} = R_{\rm sp,75\%}/R_{\rm 200m}$ (and similarly for other percentile definitions, e.g. $X_{\rm Rsp,50\%}$).  Our fiducial definition of the mass accretion rate (MAR) follows \citet{diemer_17_sparta},
\begin{equation}
    \Gamma_{\rm 200m,Tdyn}(t) = \frac{\log [M_{\rm 200m}(t)] - \log [M_{\rm 200m}(t-t_{\rm dyn,200m})]}{\log [a(t)] - \log[a(t-t_{\rm dyn,200m})]},
\end{equation}
where $a$ is the scale factor and $t_{\rm dyn}$ is the crossing time, which captures the time it takes a particle to cross a halo, $t_{\rm dyn,200m} = 2R_{\rm 200m} / V_{\rm 200m}$, where $V_{\rm 200m} = \sqrt{GM_{\rm 200m}/R_{\rm 200m}}$. We note that $t_{\rm dyn,200m}$ is independent of the halo radius and only depends on redshift and cosmology. We also investigate a number of other definitions of the MAR that are provided by \rockstar, which uses the virial mass, $M_{\rm vir}$, instead of $\mtom$ as well as different time intervals. The latter is the more important difference because the first-order difference in the mass is divided out \citep{xhakaj_19_accrate}. In particular, \rockstar provides MARs calculated over timescales of $t_{\rm dyn,vir}$, $0.5 t_{\rm dyn,vir}$, and $100\ {\rm Myr}$, which we denote $\Gamma_{\rm vir,Tdyn}$, $\Gamma_{\rm vir,0.5Tdyn}$ and $\Gamma_{\rm vir,Inst}$ respectively. We note that in the nomenclature of the \rockstar catalogues, a dynamical time is defined as a half-crossing time.

When quantifying halo mass, we not only wish to use a dimensionless quantity but also to remove the overall increase in halo mass with cosmic time. For this purpose, we use the peak height corresponding to $M_{\rm 200m}$,
\begin{equation}
    \nu \equiv \nu_{\rm 200m} = \frac{\delta_{\rm c}}{\sigma(M_{\rm 200m},z=0)D_{+}(z)},
\end{equation}
where $\delta_{\rm c}=1.686$ is the critical overdensity from the spherical top-hat collapse model \citep{Gunn72} and $D_{+}(z)$ the linear growth factor normalised to unity at $z=0$.
$\sigma$ is the RMS density fluctuation within a sphere of radius R,
\begin{equation}
    \sigma^2(M) = \sigma^2(R[M]) = \frac{1}{2\pi^2}\int_{0}^{\infty}k^2 P(k)|\Tilde{W}(kR)|^2 dk,
\end{equation}
where $M = (4\pi/3)\rhom(z=0)R^3$.
Peak heights are calculated using the \textsc{Colossus} package \citep{Colossus}.

We additionally consider the following quantities output by \rockstar: concentration, $c_{\rm 200m} \equiv \rtom / \rs$, where $\rs$ is the scale radius determined from an NFW fit \citep{navarro_97}; the normalised half-mass radius $R_{\rm 0.5M} / R_{\rm 200m}$; the strongest tidal force from any nearby halo averaged over the last half dynamical time, $F_{\rm tidal,tdyn}$, in dimensionless units of $R_{\rm vir}/R_{\rm hill}$, where $R_{\rm hill} = D(M_{\rm vir}/3M_{\rm vir,neighbour})^{1/3}$ and $D$ is the distance between two haloes; the ratio of the kinetic and potential energies for bound particles, $T/|U|$; the halo spin parameter according to \citet{Bullock01}, $\lambda_{\rm Bullock}$; ellipticity $e \equiv (a-c) / (a+b+c)$, where $a$, $b$, and $c$ are the principal axes of the halo dark matter ellipsoid and $a > b > c$; the half-mass scale $a_{\rm 0.5M}$, which captures how rapidly a halo accreted the second half of its $M_{\rm vir}$; and the maximum circular velocity, $v_{\rm max}$. We refer the reader to \citet{behroozi_13_rockstar} for a detailed description of these parameters.


\section{Correlations of the splashback radius with mass accretion rate and history}
\label{sec:corr_rsp_mar}

\begin{figure*}
	\includegraphics[width=0.98\textwidth]{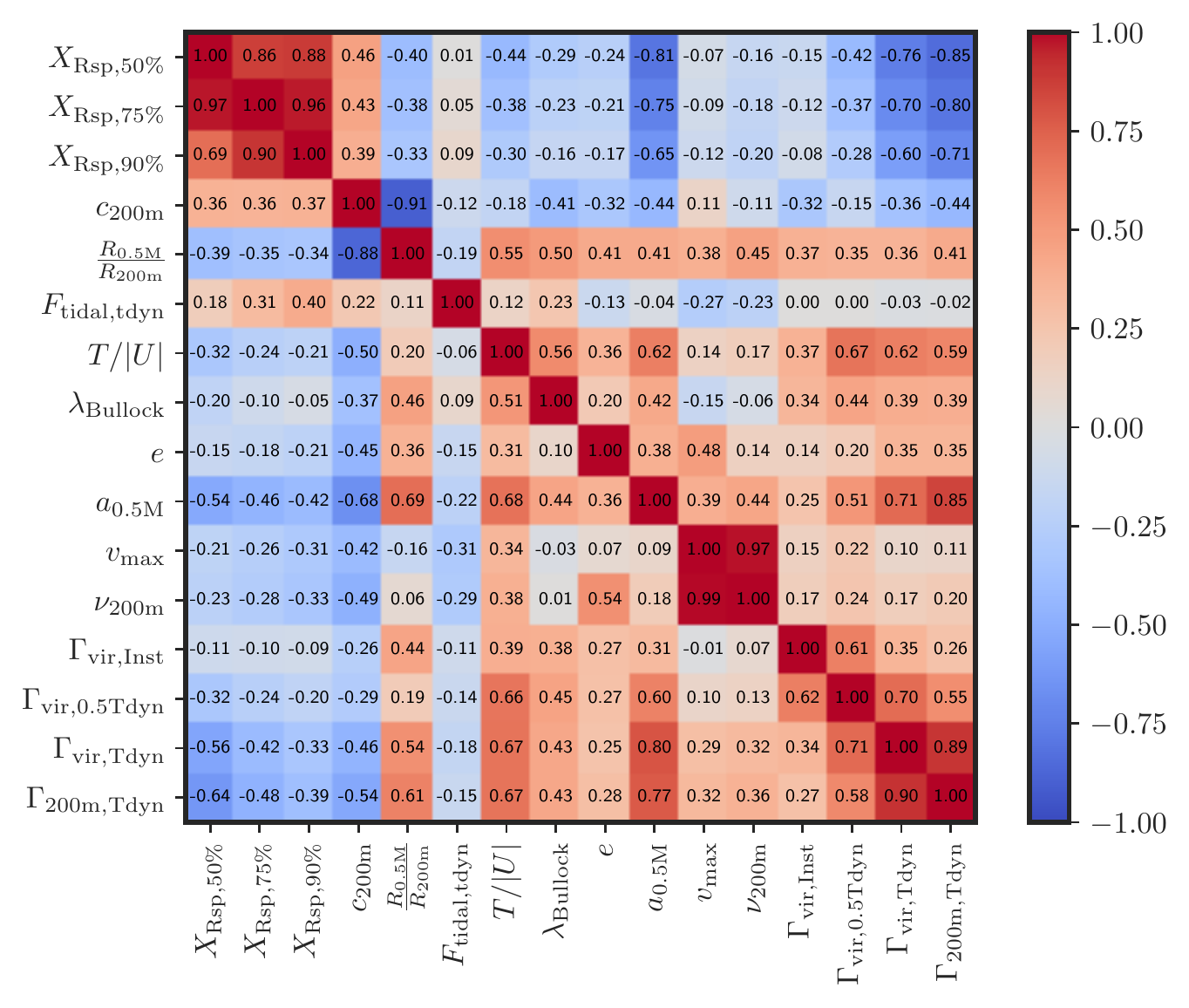}
	\caption{Pearson's correlation coefficients between halo properties at $z=0.2$ (see Section~\ref{sec:methods:defvar} for definitions).
	The lower triangle corresponds to the full sample and the upper triangle to the high-mass sample ($\nu_{\rm 200m}>2$). The trends are the same in both samples, but the correlations tend to be sharper in the high-mass sample. We observe a strong anti-correlation between the splashback radius and accretion rate, though to varying degrees for different definitions of $\Gamma$. As expected, concentration is significantly correlated with the half-mass radius, half-mass time, and accretion rate.
    }
	\label{fig:cov_mat}
\end{figure*}

In this section, we analyse the correlations of the splashback radius with a number of other halo properties (Section~\ref{sec:corr_rsp_mar:general}) and specifically focus on the well-known relation with mass accretion rate (Section~\ref{sec:corr_rsp_mar:mar}) as well as the entire accretion history (Section~\ref{sec:corr_rsp_mar:mah}). In Appendix~\ref{sec:app:pca}, we present an alternative analysis of the connection to the accretion history using principle component analysis (PCA).

We use the dimensionless quantity $X_{\rm Rsp,75\%}$ as a stand-in for the splashback radius; we show that other definitions follow very similar trends in Appendix~\ref{sec:app:rspdefs}. We restrict ourselves to two redshifts snapshots, $z=0.2$ and $z=2.0$, which represent epochs before and after dark energy becomes important, respectively. Similarly, we present two halo samples: a `full sample' of all haloes (Section~\ref{sec:methods:data_sel}) and a `high-mass sample' with haloes of peak height $\nu>2$ at $z=0.2$ (which corresponds to $M_{\rm 200m} \gsim 9 \times 10^{13} M_{\odot}$).

\subsection{General correlations between halo properties}
\label{sec:corr_rsp_mar:general}

\begin{figure*}
	\includegraphics[width=0.98\linewidth]{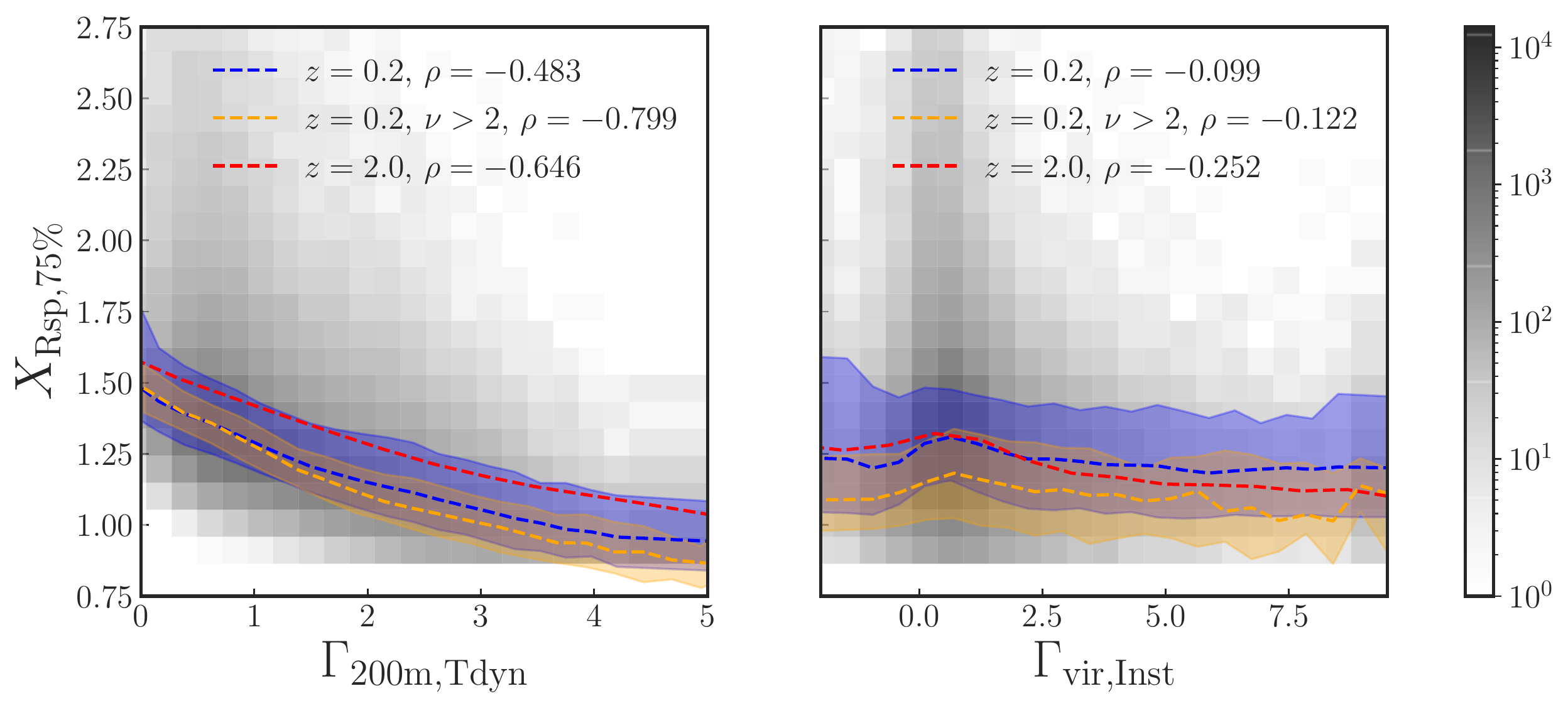}
	\caption{Relationship between $R_{\rm sp,75\%}/R_{\rm 200m}$ and accretion rate defined as $\Gamma_{\rm 200m,Tdyn}$ (left) and $\Gamma_{\rm 100Myr}$ (right). The 2D histograms show individual haloes, whereas the dashed lines and shaded areas represent the median relationships and 68\% scatter of the full, high-mass, and high-redshift samples.
	In the legend, we also show the Pearson's correlation coefficients for each sample. 
	While the accretion rate measured over one dynamical time strongly correlates with $\rsp$, the instantaneous accretion happens on a timescale too short to influence the orbital dynamics of particles.
	}
	\label{fig:rsp-acc}
\end{figure*}

Fig.~\ref{fig:cov_mat} shows the correlation between the halo properties described in Section~\ref{sec:methods:defvar} for haloes at $z=0.2$. We quantify the strength of the correlations using Pearson's coefficient,
\begin{equation}
\label{eq:corr_coef}
    \rho(x,y) = \frac{\sum_i (x_i-\Bar{x})(y_i-\Bar{y})}{\sqrt{\sum_i (x_i - \Bar{x})^2 \sum_i (y_i - \Bar{y})^2}} \,,
\end{equation}
where $x_i$ and $y_i$ represent the $i$-th data point of $x$ and $y$. Here, $\rho=1$ would indicate perfect positive correlation between $x$ and $y$ (red) whereas $\rho=-1$ would indicate perfect anti-correlation (blue). The lower triangle of Fig.~\ref{fig:cov_mat} corresponds to the full sample and and the upper triangle to the high-mass sample ($\nu_{\rm 200m}>2$). 

The normalised splashback radius ($X_{\rm Rsp}$) is highly correlated with the variables related to the accretion history, namely, $\Gamma_{\rm 200m,Tdyn}, \Gamma_{\rm vir,Tdyn}, \Gamma_{\rm vir,0.5Tdyn}$ and half-mass scale ($a_{\rm 0.5M}$). The instantaneous MAR over the last $100\ \myr$ ($\Gamma_{\rm vir,Inst}$) shows very low correlation to the splashback radius, as we discuss in detail in Section~\ref{sec:corr_rsp_mar:mar}. $X_{\rm Rsp}$ is also noticeably correlated with concentration-related properties ($c_{\rm 200m}$ and $R_{\rm 0.5M}/R_{\rm 200m}$) as well as $T/|U|$. These properties are tightly associated with the halo assembly history, as evidenced by their correlation to the MAR. The splashback radius is also moderately correlated with halo mass ($v_{\rm max}$ and $\nu_{\rm 200m}$), but in Section~\ref{sec:partial}, we show that most of this correlation is driven by the underlying correlation between mass an MAR, where higher-mass haloes accrete faster today. Interestingly, the splashback radius somewhat correlates with the tidal force around haloes ($F_{\rm tidal,tdyn}$), especially at low mass. The spin parameter ($\lambda_{\rm Bullock}$) and the ellipticity ($e$) also show interesting correlations, where fast-accreting haloes have higher angular momentum and are more elliptical. We discuss the residual impact of these properties on the splashback radius in Section~\ref{sec:partial}. 

In summary, at first sight $X_{\rm Rsp}$ correlates with all halo properties to various degrees. However, given their covariances with the MAR, these correlations may not be physically meaningful. Before removing the primary correlation, we now test which definition of the MAR is most predictive of $X_{\rm Rsp}$.

\subsection{Which definition of the accretion rate is most predictive?}
\label{sec:corr_rsp_mar:mar}

\begin{figure}
	\includegraphics[width=0.98\linewidth]{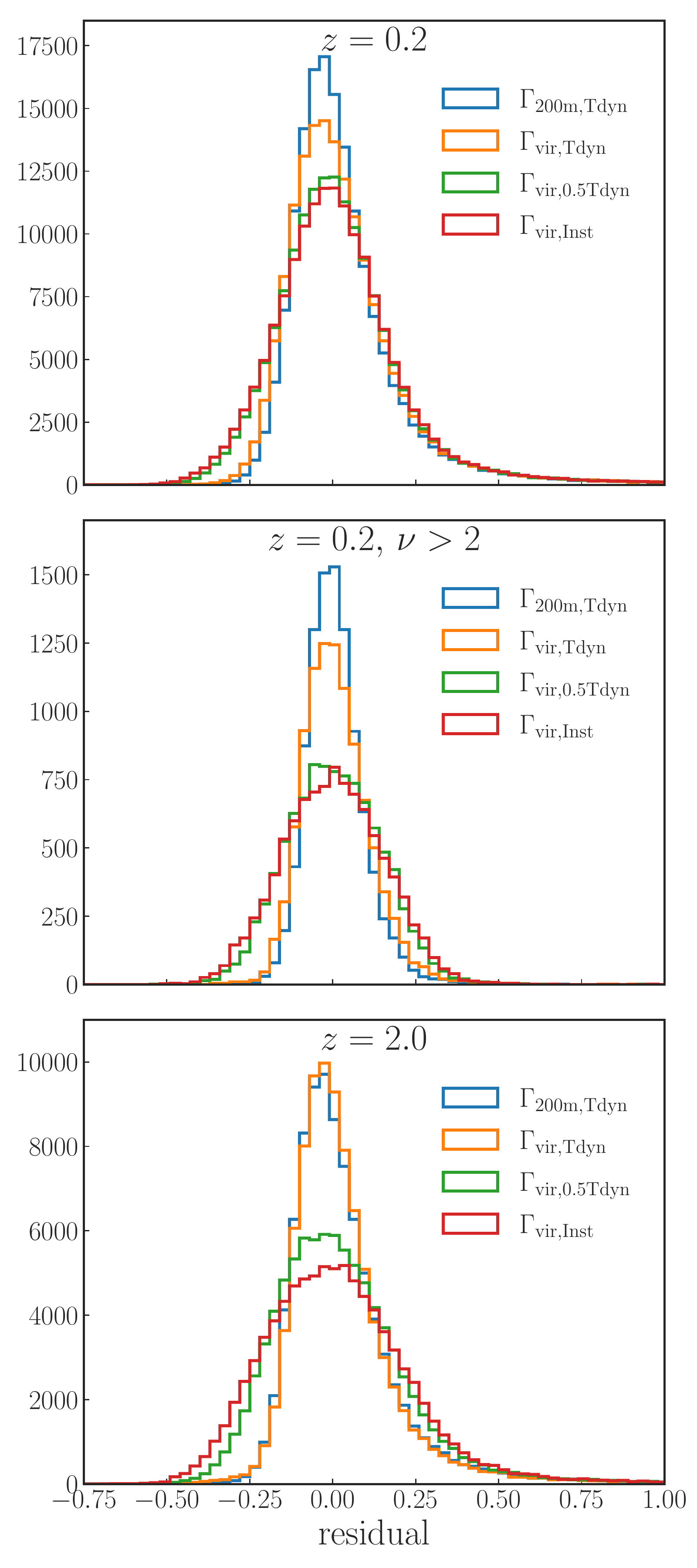}
	\caption{Histograms of the residual of $R_{\rm sp,75\%}/R_{\rm 200m}$ with respect to its median relation to different definitions of the accretion rate at $z=0.2$ (top), for high-mass haloes (middle), and at high redshift (bottom). $\Gamma_{\rm 200m,Tdyn}$ is the most predictive definition in all cases.
	}
	\label{fig:rsp-resid-defacc}
\end{figure}

When the splashback radius and its connection to MAR were first explored, it was not a priori clear which definition of the MAR would be most physically meaningful. In theoretical models with a fixed accretion rate, there is no ambiguity \citep{Adhikari14, shi_16_rsp}, but in simulations we need to measure the halo growth over some time. The crossing time is a natural choice, since that is the timescale over which particles move to their first apocenter after infall. Indeed, \citet{DK14} found that the MAR measured over approximately one crossing time was predictive of the density profiles, but the actual distribution of particles' orbiting times is complex \citep{diemer_17_sparta}, and the optimal timescale has yet to be investigated in detail. Moreover, we can vary the spherical overdensity mass definition used, although this choice has a smaller impact \citep{xhakaj_19_accrate}.

To visualise the relationship between the splashback radius and different definitions of the accretion rate, Fig.~\ref{fig:rsp-acc} shows $X_{\rm Rsp,75\%}$ as a function of $\Gamma_{\rm 200m,Tdyn}$ (the change in $\mtom$ over one halo crossing time) and the instantaneous rate measured over $100\ \myr$. 
As in all previous studies of the splashback radius, we find that the $X_{\rm sp}$ is a decreasing function of MAR (blue line and shaded area) and larger at higher redshift ($z=2.0$, red line; e.g., \citealt{MDK15}). The correlation is also tighter at high $z$ ($\rho=-0.646$ vs. $-0.483$), but that is partly due to the fixed mass limit of $1000$ particles, which means that the average peak height in the high-redshift sample is higher. As evident from the yellow line and shaded area, the correlation for $\nu > 2$ at low redshift is even stronger, $\rho=-0.799$. We observe a slight trend with mass, where $\rsp$ is slightly smaller for higher $\nu$ at fixed $\Gamma$ (in agreement with \citealt{diemer_17_rsp}). We further explore this secondary correlation in Section~\ref{sec:partial}.

The right panel of Fig.~\ref{fig:rsp-acc} contrasts these results with the `instantaneous' MAR measured over the past $100\ \myr$, $\Gamma_{\rm vir,Inst}$. Given that the crossing time is about $5\ \gyr$ at $z \approx 0$, it is not surprising that this definition of the MAR is essentially uncorrelated with the splashback radius. However, what about intermediate time intervals? To quantify the tightness of the $\Gamma$-$\xsp$ relation,  Fig.~\ref{fig:rsp-resid-defacc} shows histograms of the residuals of $X_{\rm Rsp,75\%}$ with respect to the median relationships to four different definitions of the MAR (see Section~\ref{sec:methods:defvar}). The panels show the full sample (top), the high-mass sample (middle), and the high-redshift sample (bottom). We confirm that $\Gamma_{\rm 200m,Tdyn}$ is most tightly correlated to $\xsp$ (blue histograms). Increasing the spherical overdensity threshold to $\mvir$ (and thus shrinking the radius $\rvir$ compared to $\rtom$) slightly increases the scatter (orange). Physically, this difference indicates that the mass at large radii, $\rvir < r < \rtom$, does influence the particle orbits and their apocentres. At $z \approx 2$ the definitions become essentially identical, as $\mvir \approx M_{\rm 180m}$. 

Decreasing the time interval has a larger effect on the scatter, which grows to almost the same level as for $\Gamma_{\rm vir,Inst}$ (red) when halving the interval (green). Physically, this strong difference in scatter indicates that accretion during the first half-crossing time (from infall to pericentre) matters as much for the energetics of the particle orbits as accretion during their second half-orbit (from pericentre to apocentre). The effect of the time interval is even more pronounced in the high-mass sample. In the full sample, all definitions experience roughly equal tails towards small and large residuals, indicating that those are not caused by the MAR but by halo physics or by noise in the determination of $\rsp$ (see also Figs.~\ref{fig:cov_mat} and \ref{fig:rsp-acc}).

In summary, we confirm a picture where splashback is most correlated with an accretion rate measured over one crossing time, and where the halo mass should be considered to a relatively large radius ($\rtom$). However, we have not probed even longer time intervals in the past, or whether it makes sense to summarise the accretion history in a single number. We thus turn to the correlation of $\xsp$ with the entire history now.

\subsection{Which periods in the accretion history matter most?}
\label{sec:corr_rsp_mar:mah}

\begin{figure}
	\includegraphics[width=0.98\linewidth]{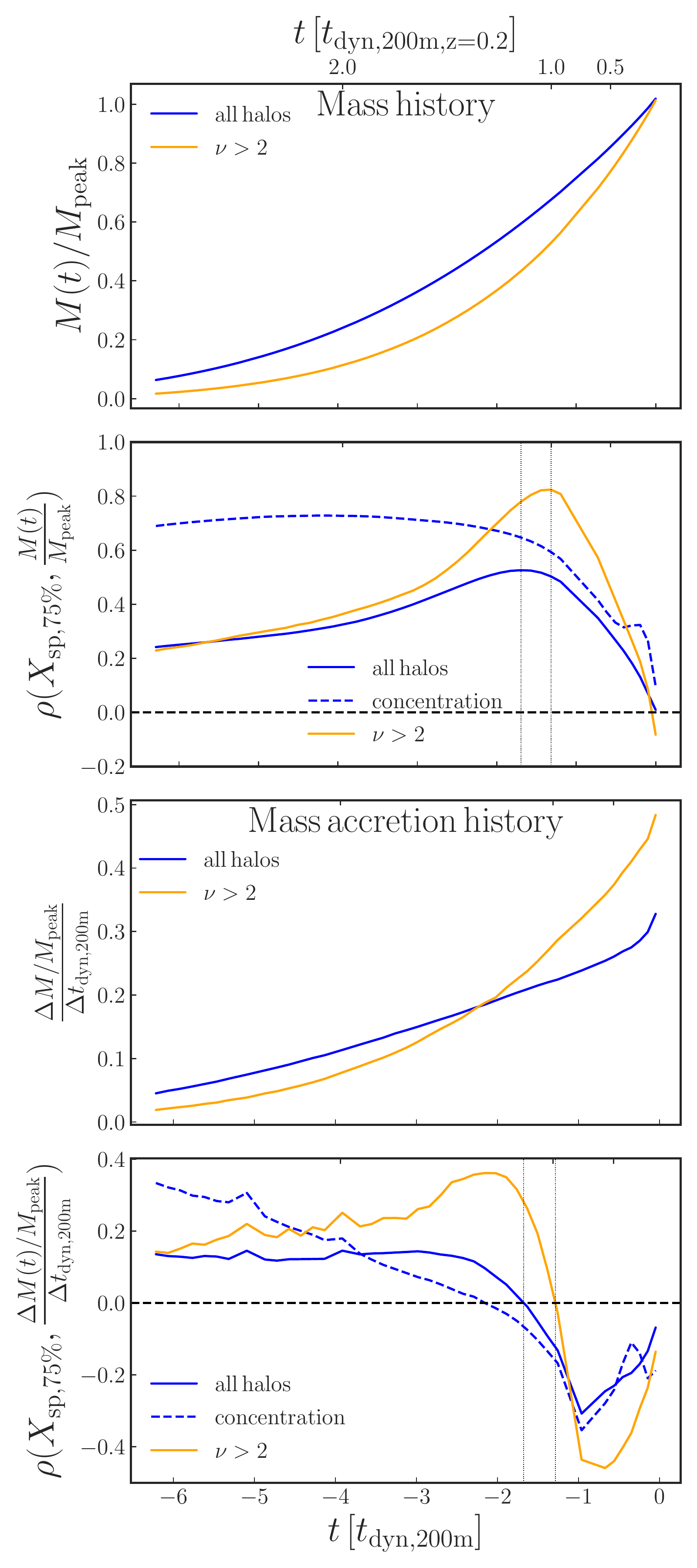}
	\caption{Cumulative mass history (top panel) and differential mass accretion history (third panel) of all (blue) and massive (orange) haloes, and how they are correlated to $\rsp$ (second and fourth panels). The $x$-axes show dynamical times in the past, using both the integrated dynamical time (bottom) and current dynamical time at $t = 0$ (top). The average $M_{\rm 200m}$ history is normalised by the peak mass, and the splashback correlation peaks roughly one dynamical time ago (vertical dotted lines in second panel), meaning haloes that have grown little since have a large $\rsp$. The correlation to the accretion rate history (bottom panel) is easier to interpret: low rates over the last dynamical time mean a large $\rsp$, and vice versa (the dotted lines now mark the zero crossing of the correlation). At much earlier times, the correlation becomes more or less flat. The dashed blue lines show the correlation with concentration for comparison, which is more sensitive to early epochs (the formation time). 
	}
	\label{fig:M_history}
\end{figure}

Inspired by the results of the previous section, we use the dynamical time, $t_{\rm dyn,200m}$, as our unit of time (Section~\ref{sec:methods:defvar}). The number of dynamical times between two redshifts $z_1$ and $z_2$ (with $z_1>z_2$) is calculated as
\begin{equation}
    t[t_{\rm dyn,200m}]=\int_{z_1}^{z_2} dz \frac{dt/dz}{t_{\rm dyn,200m}(z)} \,.
\end{equation}
We also consider the \textit{current} dynamical time at $z = 0.2$, $t[t_{\rm dyn,200m,z=0.2}]$, which is roughly $\sim 4 \gyr$. This definition is used to calculate $\Gamma_{\rm 200m,Tdyn}$. We calculate individual mass accretion histories by normalising $M_{\rm 200m}$ of haloes at each redshift by their maximum mass at any epoch, $M_{\rm peak}$. In addition to this `mass history,' we also consider the `accretion history' $(\Delta M / M_{\rm peak})/\Delta t_{\rm dyn,200m}$, the change per time of $M_{\rm 200m} / M_{\rm peak}$ between adjacent snapshots.

Fig.~\ref{fig:M_history} shows the mass and accretion histories from $z = 0.2$ (or $t = 0$) to $z = 4$ (top and third panel, respectively). Naturally, haloes gain mass on average. As expected in hierarchical structure formation, the high-mass sample (orange) forms more recently, on average accreting half of its current mass over the last $\sim$1.5 dynamical times. This recent growth is mirrored in the third panel, where we see the accretion rate reach up to half the peak mass per dynamical time at $z \approx 0$.

In the second panel of Fig.~\ref{fig:M_history}, we quantify the correlation between $X_{\rm Rsp,75\%}$ and each epoch in the mass history using Pearson's correlation coefficient. At first sight, it may seem counter-intuitive that this correlation is positive at all epochs, but a higher (normalised) mass at any epoch in the past means a lower mass accretion rate today, and thus a larger $\xsp$. Interestingly, the correlation peaks $\sim$1--2 dynamical times in the past, or about one current dynamical time (upper ticks), indicating that the accretion rate thereafter has a particularly strong impact on $\xsp$. The correlation of the high-mass sample (orange) peaks slightly later than for the full sample. While it is not clear whether this difference is statistically significant, we speculate that the spatial extent of the halo that really matters for particle dynamics is the splashback radius, which is slightly smaller on average for the faster-accreting high-mass sample. On the other hand, we cannot exclude other effects, e.g., from the haloes' environments.

These arguments become easier to understand when considering the correlation between $\xsp$ and the accretion history in the bottom panel of Fig.~\ref{fig:M_history}. The correlation is more or less flat at early times and turns negative between $1$ and $2$ dynamical times (or one current dynamical time) in the past. The peak in the correlation with the mass history corresponds to the zero-crossing of the accretion history correlations (dotted vertical lines). The rapid switch is particularly apparent in the high-mass sample, once again supporting our picture where the accretion during particles' first orbit from infall to apocentre, but not before infall, sets $\xsp$. We also observe a weakening of the correlation towards $t = 0$, which indicates that accretion during the very last phase of the particle orbits comes too late to significantly influence their apocentres. It is unclear whether there is a fundamental mass dependence to these trends, or whether the high-mass sample is simply less affected by neighbouring haloes and stochastic noise in the particle orbits. Our results for the high-mass sample show that measuring the MAR over one current dynamical time is, indeed, a near-optimal estimator of the impact of the MAH on $\rsp$. In low-mass haloes, $\rsp$ tends to be larger than $\rtom$, meaning that infalling particles have to travel slightly longer distances than implied by the $\rtom$ definition of the MAR.

For comparison, we also consider the correlations of concentration with the mass and accretion histories (dashed lines in Fig.~\ref{fig:M_history}). Concentration has long been known to be tightly connected to the formation time of haloes \citep{navarro_97, Bullock01, wechsler_02, ludlow_13}. Naturally, formation time (which can be captured by the half-mass time, $a_{\rm 0.5M}$) and current accretion rate are strongly correlated ($\rho=0.85$ for the high-mass sample in Fig.\ref{fig:cov_mat}), but it has recently been shown that the density profiles of haloes retain separate information from these different epochs: the formation time sets the concentration, whereas the MAR affects the entire profile and particularly the position of the splashback radius \citep{luciesmith_22_mah, diemer_22_prof1}. Fig.~\ref{fig:M_history} confirms that the correlation between $c_{\rm 200m}$ and accretion history (bottom panel) does not flatten like that of $\xsp$ but instead rises monotonically going back in time (see Fig~1 of \citet{Wang20} for a similar result). As a consequence, early-forming haloes have high concentrations, as expected. At very recent times, the correlation inverts, presumably because major mergers tend to temporarily increase the concentration \citep{Wang20}.

In summary, we have, for the first time, analysed how the splashback radius is set by different epochs in haloes' accretion histories, and confirmed that the late-time accretion over one crossing time is the key determinant. One caveat is that we have used averaged accretion histories and their correlations, leaving open the question of whether there are outlier haloes. In Appendix~\ref{sec:app:pca}, we further explore the history-splashback connection for individual haloes using a PCA technique.


\section{Correlations with other halo properties}
\label{sec:partial}

\begin{figure*}
	\includegraphics[width=0.98\textwidth]{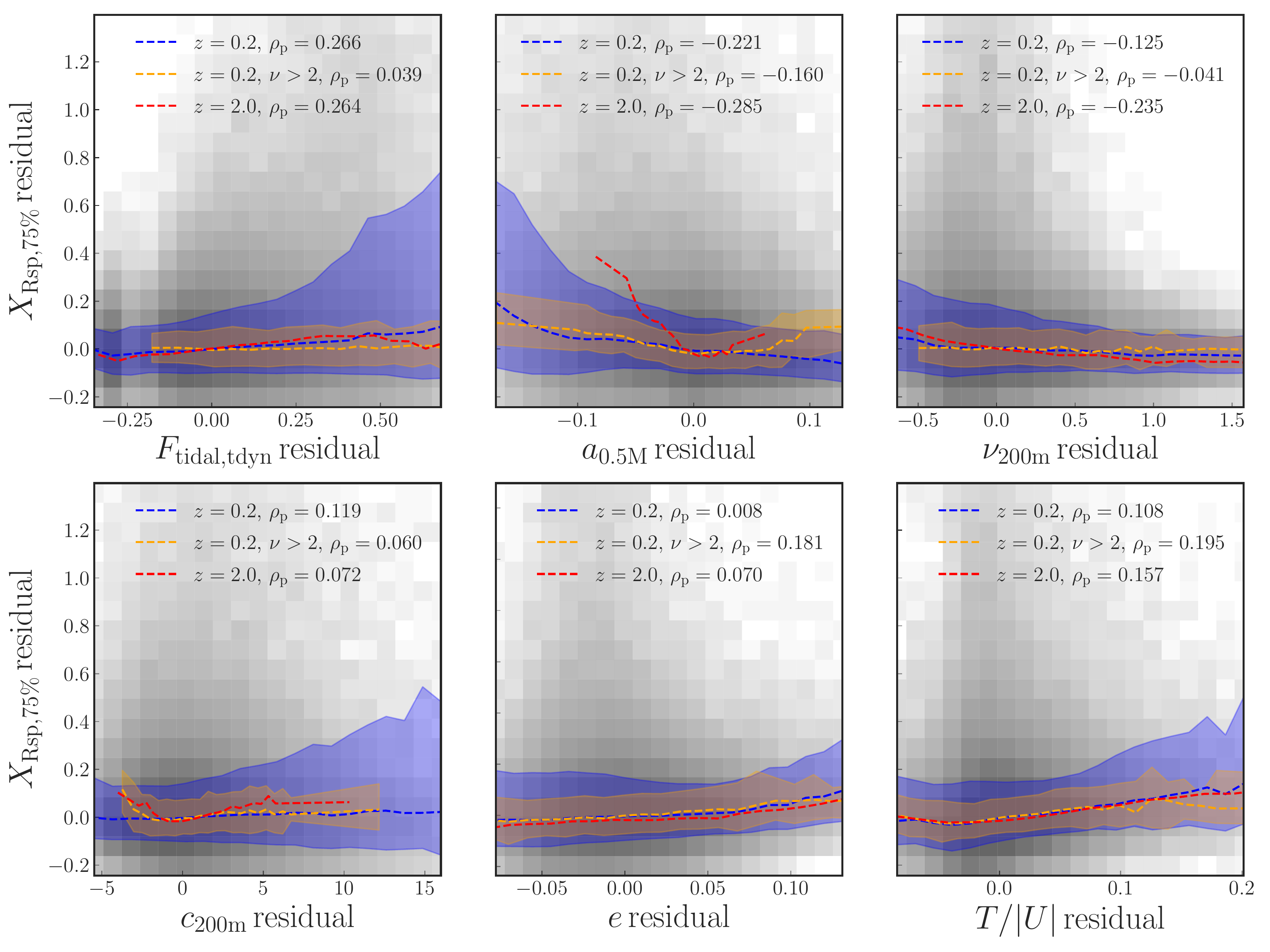}
	\caption{Secondary correlations between halo properties and the residual in the splashback radius after the primary correlation with MAR has been removed. As in Fig.~\ref{fig:rsp-resid-defacc}, the residuals are computed with respect to the median $X_{\rm sp,75\%} = R_{\rm sp,75\%}/R_{\rm 200m}$ at fixed $\Gamma_{\rm 200m,Tdyn}$. The 2D histograms show individual haloes in the full sample, whereas the lines and shaded areas show the median relations and 68\% scatter. The partial correlation coefficients shown in the legend are much smaller than those with mass accretion rate in Fig.~\ref{fig:rsp-acc}, highlighting that we find no strong secondary correlations with any variable.
	}
\label{fig:partial_rsp}
\end{figure*}

\begin{figure}
	\includegraphics[width=0.98\linewidth]{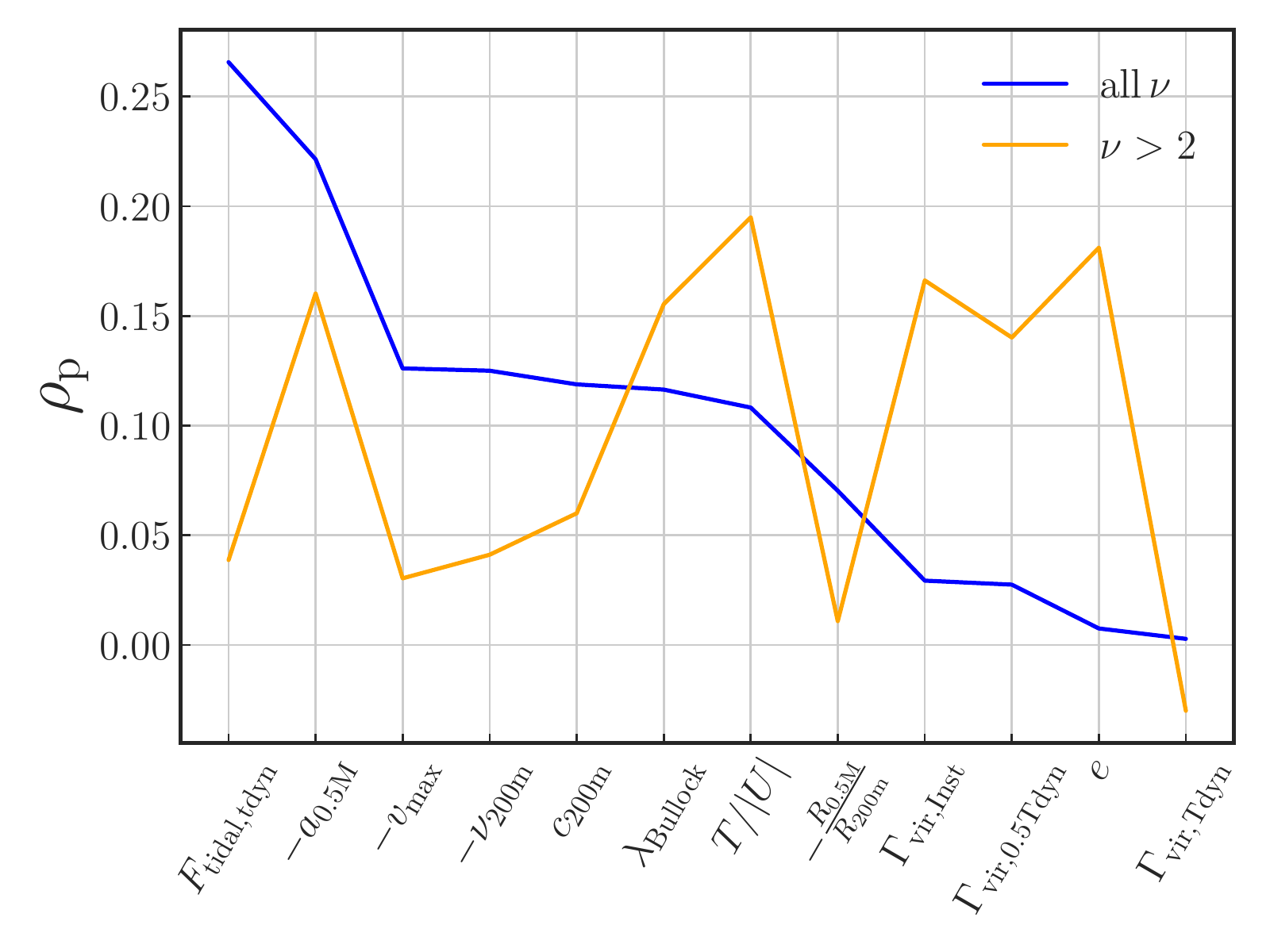}
	\caption{Partial correlation coefficients between various halo properties and the residuals of $X_{\rm sp,75\%} = R_{\rm sp,75\%}/R_{\rm 200m}$ with respect to its median relation with $\Gamma_{\rm 200m,Tdyn}$. Minus signs indicate halo properties where $\rho_\rmp$ is negative, i.e., where a higher halo property means lower $\xsp$. The behaviour at high redshift ($z = 2$) is similar to the $z=0.2$ sample, but the high-mass sample (orange) exhibits somewhat different correlation coefficients.
    }
	\label{fig:corr_chart}
\end{figure}

We have established that $\rsp$ is primarily determined by the MAR, but Fig.~\ref{fig:cov_mat} showed a number of other, potentially interesting correlations. In this section, we test whether these correlations are `real' or purely driven by covariances with the MAR. Specifically, Fig.~\ref{fig:partial_rsp} is laid out like Fig.~\ref{fig:rsp-acc}, but instead of the correlations between $\xsp$ and MAR it shows secondary (or partial) correlations with other variables. Here, we have subtracted the median $\xsp$ at the $\Gamma_{\rm 200m,Tdyn}$ of each halo, i.e., the dominant correlation with MAR shown in Fig.~\ref{fig:rsp-acc}. Based on the residual relation, we also compute the partial Pearson's correlation coefficient, $\rho_{\rm p} (x,X_{\rm sp}|\Gamma)$, between the residual of $\xsp$ and another variable $x$.

Our first impression is that the secondary correlations are weak at best: the median relations (dashed lines) are more or less flat, and Pearson's correlation coefficients are low. This conclusion also holds for the other halo properties that are not shown in Fig.~\ref{fig:partial_rsp}, which we summarise in Fig.~\ref{fig:corr_chart}. Here, we have ordered the properties by the strength of their correlation for the full sample (blue). The high-mass sample exhibits somewhat different values (orange). The six properties shown in Fig.~\ref{fig:partial_rsp} were selected either because they are important in general or because they have an interesting, albeit small, partial correlation with $\xsp$. We now discuss these six halo properties in more detail (Sections~\ref{sec:partial:tidal}--\ref{sec:partial:ellipticity}), and explore whether the magnitude gap could be a relevant observable (Section~\ref{sec:partial:mgap}).

\subsection{Tidal force}
\label{sec:partial:tidal}

We have already speculated that the environment of haloes might affect their splashback radius. Since we are most interested in forces acting on particles' orbits, the tidal force from surrounding haloes might be a good way to capture these effects. Specifically, $F_{\rm tidal,tdyn}$ represents the strongest tidal force from any nearby halo averaged over the last 0.5$t_{\rm dyn,vir}$ in dimensionless units. The top-left panel of Fig.~\ref{fig:partial_rsp} shows that there is a positive correlation ($\rho_{\rm p}=0.266$), which would be compatible with a picture where particles experience outward forces that drive them to larger apocentres. The correlation seems to be most significant at the strongest tidal forces, and it vanishes for high-mass haloes (orange line), which tend to dominate their environment and are thus less susceptible to tidal influences. The correlation persists at higher redshift (red line).

\subsection{Half-mass scale and concentration}
\label{sec:partial:halfmass}

As discussed in Section~\ref{sec:corr_rsp_mar:mah}, the half-mass scale $a_{\rm 0.5M}$ captures the general age of a halo as opposed to its late-time accretion, although the two are strongly correlated (Fig.~\ref{fig:cov_mat}). The age of haloes, in turn, affects their clustering, an effect known as `assembly bias' \citep{gao_05_ab}. We could thus imagine a number of mechanisms by which halo age might affect the splashback radius. The top-middle panel of Fig.~\ref{fig:partial_rsp} supports the picture from Fig.~\ref{fig:M_history}, where $\xsp$ is almost exclusively set by late-time accretion but is slightly sensitive to halo growth at earlier times. The resulting trend is a slight negative correlation, where earlier-forming haloes have higher $\xsp$ at fixed $\Gamma$. However, the scatter grows towards early $a_{\rm 0.5M}$ because there is a longer time between the formation time and the last dynamical time, during which the haloes' histories have time to diverge. The high-mass haloes (orange) with negative residuals show a similar trend as the full sample, but that trend reverses for late $a_{\rm 0.5M}$. We speculate that this counter-intuitive trend might be due to frequent major mergers in the recent past of very late-forming, massive haloes. 

Given that the half-mass scale reliably predicts concentration (Section~\ref{sec:corr_rsp_mar:mah}), we might expect that the mild secondary correlation of $\xsp$ and $a_{\rm 0.5M}$ would translate into a similar correlation with $c_{\rm 200m}$. However, the bottom-left panel of Fig.~\ref{fig:partial_rsp} indicates a non-significant correlation, meaning that almost the entire correlation observed in Fig.~\ref{fig:cov_mat} is a product of the relationship between concentration and MAR.

\subsection{Halo mass (peak height)}
\label{sec:partial:peakheight}

Since gravity is intrinsically scale-free, the dynamics of haloes, and thus their $\xsp$, should not change with halo mass. However, previous work found that $\xsp$ is slightly smaller at fixed $\Gamma$ for higher-mass haloes \citep{diemer_17_rsp, diemer_20_catalogs}. We confirm this picture in the top-right panel of Fig.~\ref{fig:partial_rsp}, where the full sample shows a weak anti-correlation ($\rho_{\rm p}=-0.125$) that appears to be caused by low-mass haloes (negative residuals). This trend is even slightly stronger at $z = 2$. The correlation almost vanishes for the high-mass sample. While surprising at first, this result seems in rough agreement with Fig. 3 in \citet{diemer_17_rsp}, which shows that the mass trend becomes very weak at $\nu > 2$ and $z = 0$.

\subsection{Ellipticity and kinetic energy}
\label{sec:partial:ellipticity}

One uncertainty in the dynamical determination is the large spread of particle apocentres, and non-sphericity certainly contributes a significant fraction of that scatter \citep{diemer_17_sparta}. We could thus easily imagine that making a halo more elliptical would mean to distribute a part of its particles' apocentres to larger radii, and thus to increase high percentiles of the distribution (such as the 75th percentile used here). While the formal correlation coefficient between $\xsp$ and $e$ in the bottom-centre panel of Fig.~\ref{fig:partial_rsp} is low, the median relation does exhibit such a trend. Moreover, in the high-mass sample the partial correlation coefficient is $0.181$, comparable to those analysed in the previous sections. This result highlights why it is important to consider secondary correlations: overall, ellipticity is quite strongly anti-correlated with $\xsp$ (Fig.~\ref{fig:cov_mat}) because a high $e$ goes hand in hand with high accretion rate, and thus a low $\xsp$.

Another quantity that correlates strongly with both MAR and splashback radius is $T/|U|$, the ratio of the kinetic and potential energies of particles in a halo. As for ellipticity, however, the bottom-right panel of Fig.~\ref{fig:partial_rsp} shows that the secondary correlation with $\xsp$ is positive, which we would expect given that higher kinetic energy at fixed potential energy (fixed mass) should mean larger orbits and thus larger apocentres. 

\subsection{The magnitude gap}
\label{sec:partial:mgap}

\begin{figure}
	\includegraphics[width=0.98\linewidth]{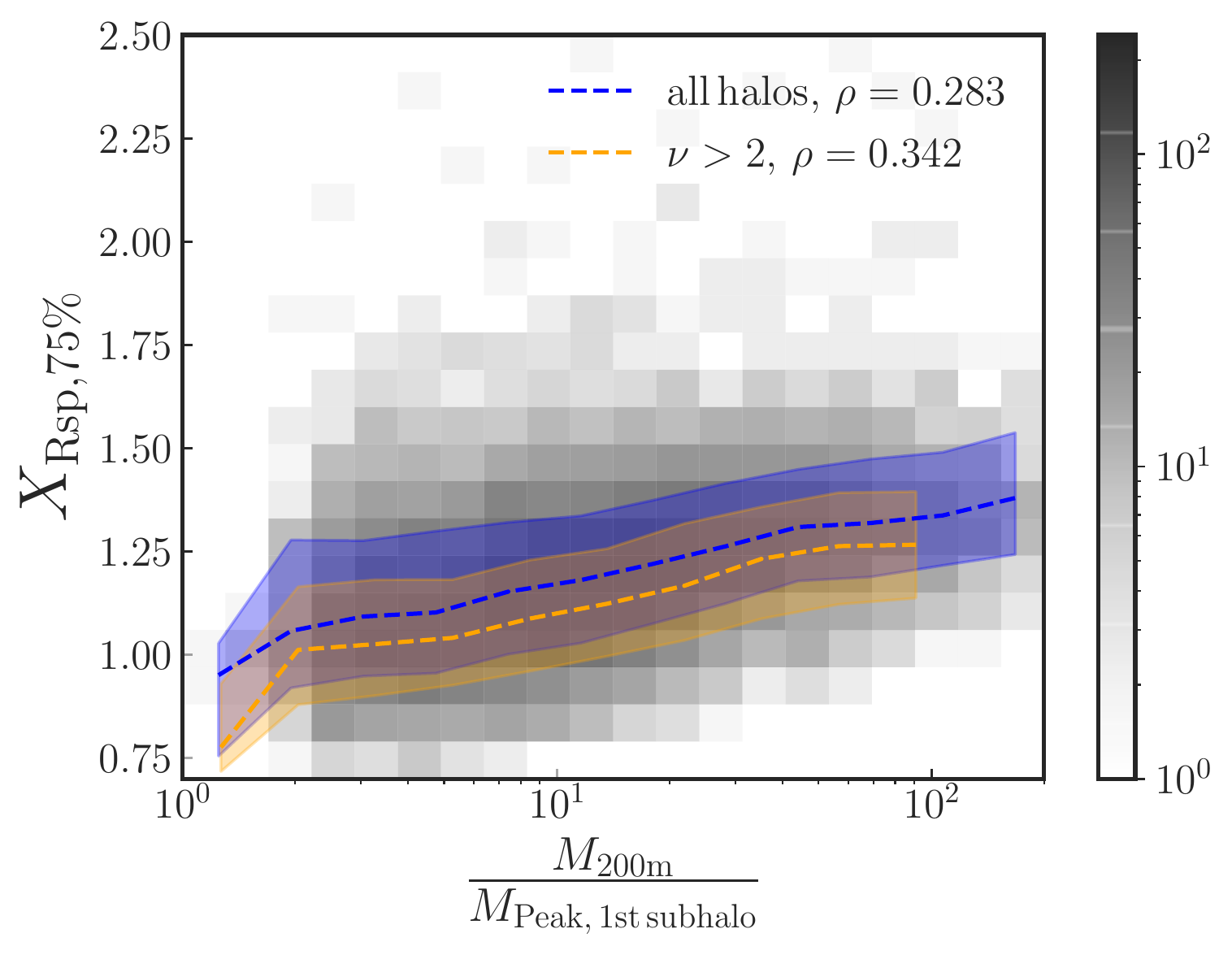}
	\caption{Relationship between the splashback radius, $R_{\rm sp,75\%}/R_{\rm 200m}$, and the mass gap (the ratio of the parent's $M_{\rm 200m}$ to the most massive subhalo's $M_{\rm peak}$). This quantity mimics the observable magnitude gap between the the central galaxy and the brightest (or $n$-th brightest) satellite. We find the expected positive correlation, where the magnitude gap grows with age and thus predicts a higher $\xsp$.
	}
	\label{fig:mass_gap}
\end{figure}

While the correlations we have discussed in the previous sections tell us about the physics of halo formation, the respective quantities are generally not directly observable, at least not for individual haloes. One possible indicator of halo growth that is observable is the `magnitude gap,' the difference in magnitude between the brightest cluster galaxy (BCG) and its $n$-th brightest satellite \citep{Jones03, Donghia05, Dariush10, hearin_13_gap, Deason13, Kundert17, GoldenMark18, Vitorelli18, Farahi20}. Satellites get stripped and tidally disrupted over time, and these processes particularly affect the largest satellites, which suffer most dynamical friction \citep{chandrasekhar_43}. As time passes since a halo actively grew and accreted massive satellites, the brightness gap between the BCG and brightest satellite thus grows on average, meaning that a large magnitude gap is an indicator of an old halo age or a low (current) accretion rate. We would thus expect the size of the magnitude gap to positively correlate with $\xsp$.

To investigate this question without artificially `pasting' galaxies on the haloes in our $N$-body simulations, we use a simple proxy for the magnitude gap: the ratio of $M_{\rm 200m}$ of the parent halo and the peak mass of the most massive subhalo. Assuming that satellite magnitudes are decreasing (being brighter) with increasing subhalo mass, this quantity qualitatively reflects the the magnitude gap. Since we need the largest subhalo to be well resolved, we apply a more stringent cut on the host halo mass, namely, 20,000, 20,000, 15,000, 10,000, and 5,000 dark matter particles for the L0063, L0125, L0250, L0500 and L1000 simulations, respectively. The corresponding halo masses are $3.4 \times 10^{11}$, $2.8 \times 10^{12}$, $1.65 \times 10^{13}$, $8.7 \times 10^{13}$ and $3.5 \times 10^{14} M_{\odot}$. We have checked that the results are converged with these limits.

Fig.~\ref{fig:mass_gap} shows that the mass ratio does exhibit the expected positive correlation with $\xsp$. The size of the correlation coefficient ($\rho = -0.28$) is comparable to many of the correlations in Fig.~\ref{fig:cov_mat}, but it should be observable even in individual haloes (with the caveat of the translation from subhaloes to galaxies). The scatter in the relation is, however, comparable to the size of the overall trend. We have confirmed that the lower $\xsp$ in the high-mass sample at fixed subhalo mass ratio is driven by the higher MARs in the high-mass sample.

In summary, given that the density profile of individual haloes is not observable with current weak lensing or satellite profiles, the magnitude gap is, to our knowledge, the most promising single observable to split haloes of fixed mass into samples that should exhibit different splashback radii. However, this connection should be checked in hydrodynamical simulations of galaxy formation, where actual magnitudes are available.


\section{Conclusion}
\label{sec:conclusion}

We have presented a thorough assessment of which properties of dark matter haloes set the (relative) position of their splashback radius. We have confirmed and refined the previous finding that the accretion history is the most important factor, but we have also discovered a number of interesting secondary effects. Our most important conclusions are as follows.

\begin{enumerate}

    \item The accretion rate over the last crossing time is by far the strongest predictor of the splashback radius, with a physically well-understood anti-correlation. Changing the mass definition from $\mtom$ to higher overdensities such as $\mvir$, or reducing the time interval over which the rate is measured, increase the scatter in the relation.

    \item Correlating the splashback radius with the full mass accretion history reveals that $\rsp$ suddenly becomes sensitive to the recent accretion at one current dynamical time in the past (or $1.5$ integrated dynamical times). As a result, $\rsp$ is most positively correlated with the cumulative, fractional mass at that time. These results support a picture where particles are influenced by mass growth only between their infall and first apocentre. 
    
    \item We confirm that halo concentration is also tightly related to the accretion history, but that it captures the overall age of the halo rather than its recent accretion.
    
    \item We study the effect of other halo properties on the splashback radius by considering their impact at fixed mass accretion rate. All secondary correlations are comparatively weak, but we confirm that more massive haloes have slightly smaller $\rsp$. An early formation time, strong tidal forces from neighbours, high ellipticity, and high kinetic-to-potential energy ratio can all slightly increase $\rsp$ at fixed accretion rate. 
    
    \item In a principal component analysis of the mass history, the first component explains 69\% of the halo-to-halo variations and strongly correlates with the splashback radius. We build a polynomial fitting function based on the first three components, but this function predicts the splashback radius only slightly more accurately than a simple relation with accretion rate. 

    
\end{enumerate}

One caveat to our analysis is that Pearson's correlation coefficient captures only the linear relationships between variables but ignores any non-linear effects. Similarly, we have neglected baryonic effects. We have focused on the particular splashback scale of haloes as determined by particle dynamics, but in future work this analysis could be extended to the full density profiles of haloes \citep[e.g.,][]{xhakaj_22}.

\section*{Acknowledgements}
TS is supported by the US Department of Energy under award DE-SC0018053. 
Many of the computations were run on the \textsc{Midway} computing cluster provided by the University of Chicago Research Computing Center and on the DeepThought2 cluster at the University of Maryland.


\bibliographystyle{mnras}
\bibliography{reference,bib_mine} 


\appendix


\section{Correlations with different splashback definitions}
\label{sec:app:rspdefs}

\begin{figure}
	\includegraphics[width=0.98\linewidth]{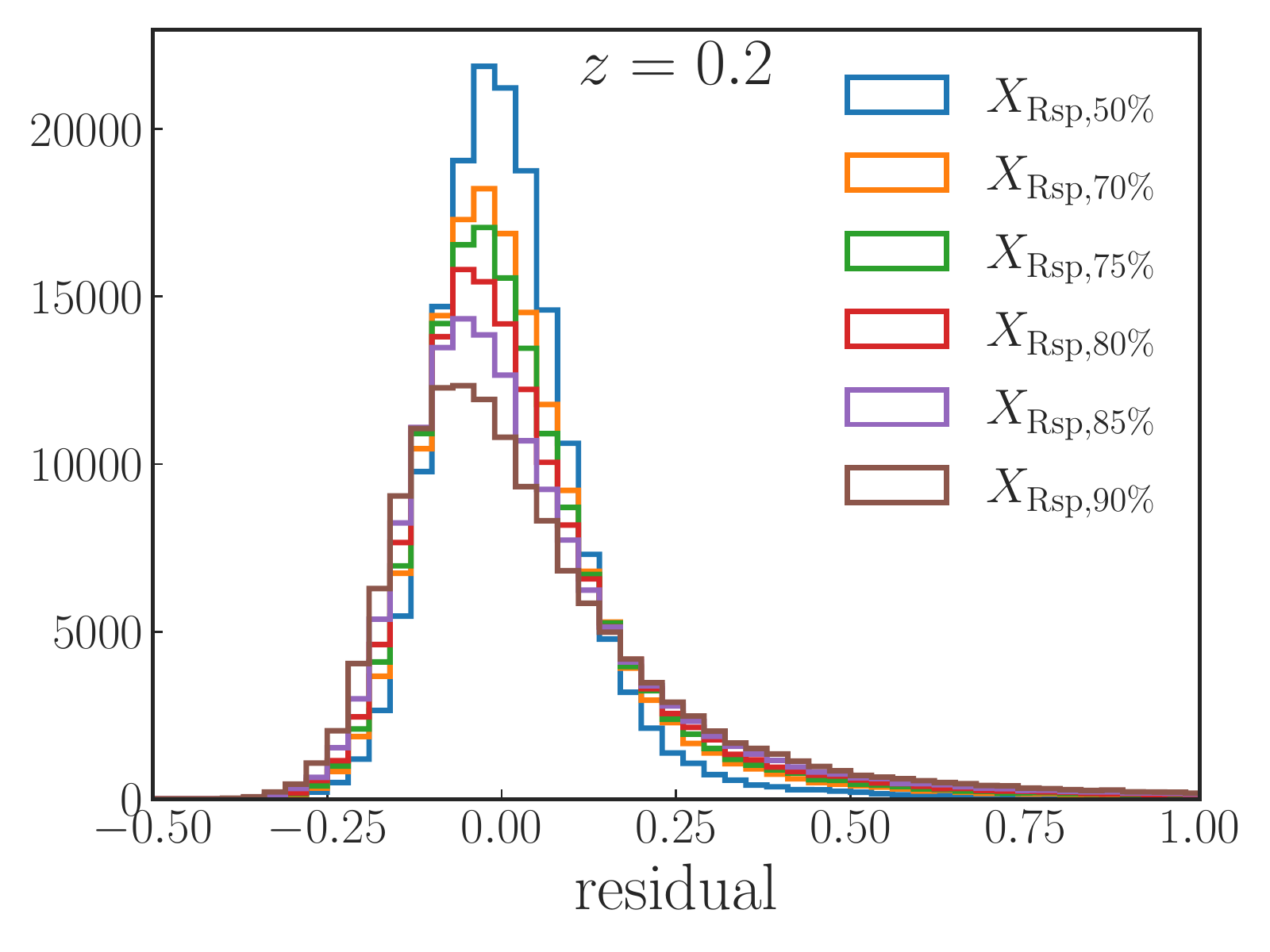}
	\caption{Distribution of the residual of different splashback definitions with rep definitions with respect to their median relationship to $\Gamma_{\rm 200m,Tdyn}$. The scatter increases for higher percentiles.
    }
	\label{fig:hist_resid_p7090}
\end{figure}

\begin{figure}
	\includegraphics[width=0.98\linewidth]{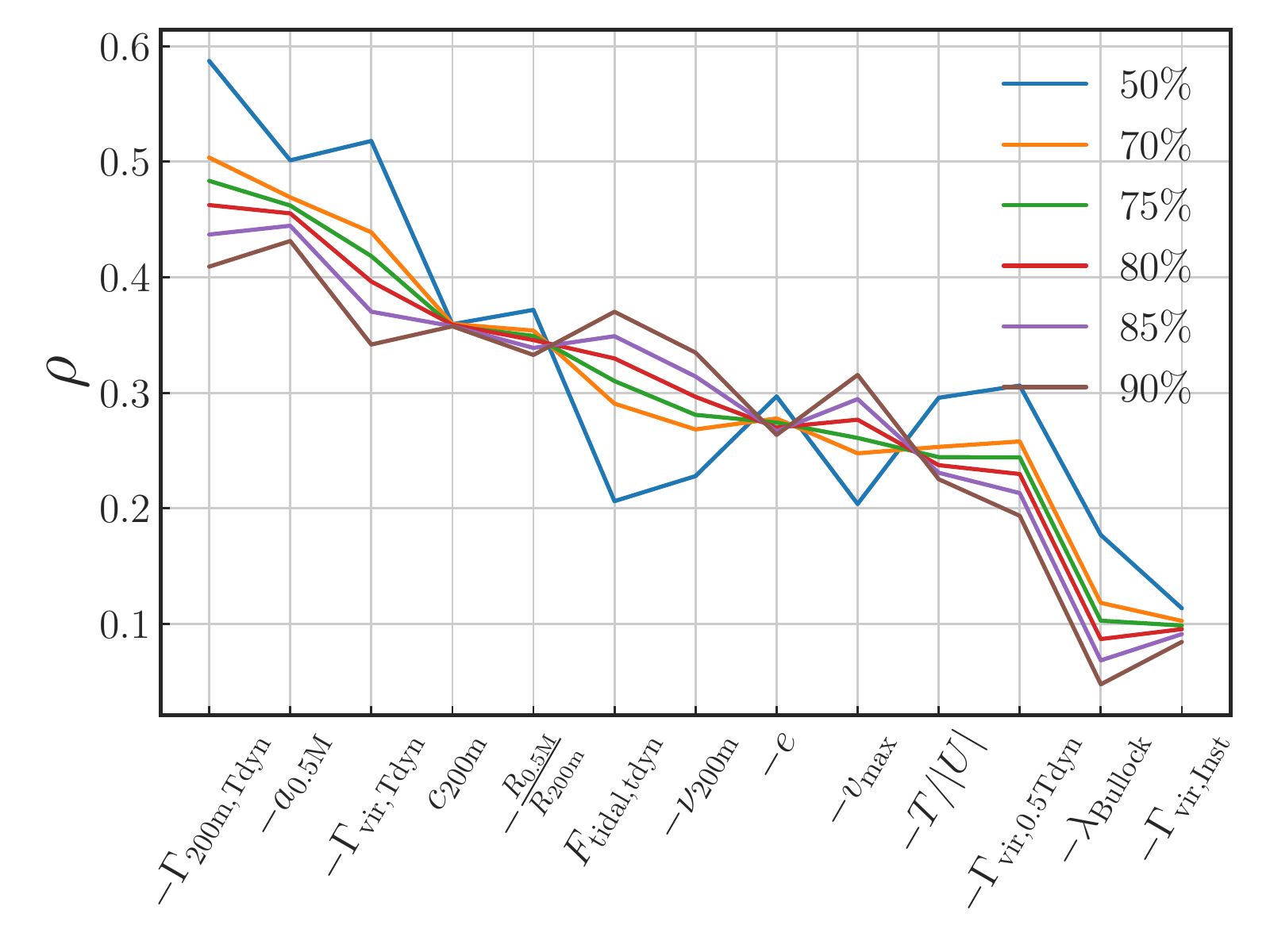}
	\caption{Pearson's correlation coefficients between the normalised splashback radius, $X_{\rm sp,P} = R_{\rm sp,P}/R_{\rm 200m}$, and various cluster properties for different definition of splashback radius, where P=[50\%, 70\%, 75\%, 80\%, 85\%, 90\%] (see Section~\ref{sec:methods:sparta}).
    }
	\label{fig:corr_chart_app}
\end{figure}

Throughout the paper, we have quantified correlations with the splashback radius as defined by the 75th percentile of particle apocentres, but our results hold for other definitions as we show in this Appendix. First, Fig.~\ref{fig:hist_resid_p7090} repeats the $z \approx 0$ panel of Fig.~\ref{fig:rsp-acc}, but it shows the residuals of different percentile definitions (Sec.~\ref{sec:methods:sparta}) over their median relationship with $\Gamma_{\rm 200m,Tdyn}$. The scatter monotonically increases with radius for two reasons. First, percentiles near the median (50th percentile) carry a smaller statistical error than those near the edge of the distribution (e.g., the 90th percentile). Secondly, and more importantly, the tail of the apocentre distribution is more sensitive to effects such as non-sphericity and environment \citep{diemer_17_sparta}. The 75th percentile represents an intermediate value with intermediate scatter, and our results for the connection between $\rsp$ and the accretion history qualitatively hold for the other definitions.

We also need to check whether the secondary correlations found in Section~\ref{sec:partial} differ systematically for different definitions of the splashback radius. We summarise the results in Fig.~\ref{fig:corr_chart_app}, which shows Pearson's correlation coefficients for different percentile definitions and various halo quantities. Overall, the correlations are very similar for different definitions, and the trends with percentile are monotonic for each halo property. Most notably, the correlation with accretion rate is even significantly stronger for the 50th percentile than for the 75th percentile that we analysed throughout the paper.


\section{Principal component analysis}
\label{sec:app:pca}

\begin{figure}
	\includegraphics[width=0.98\linewidth]{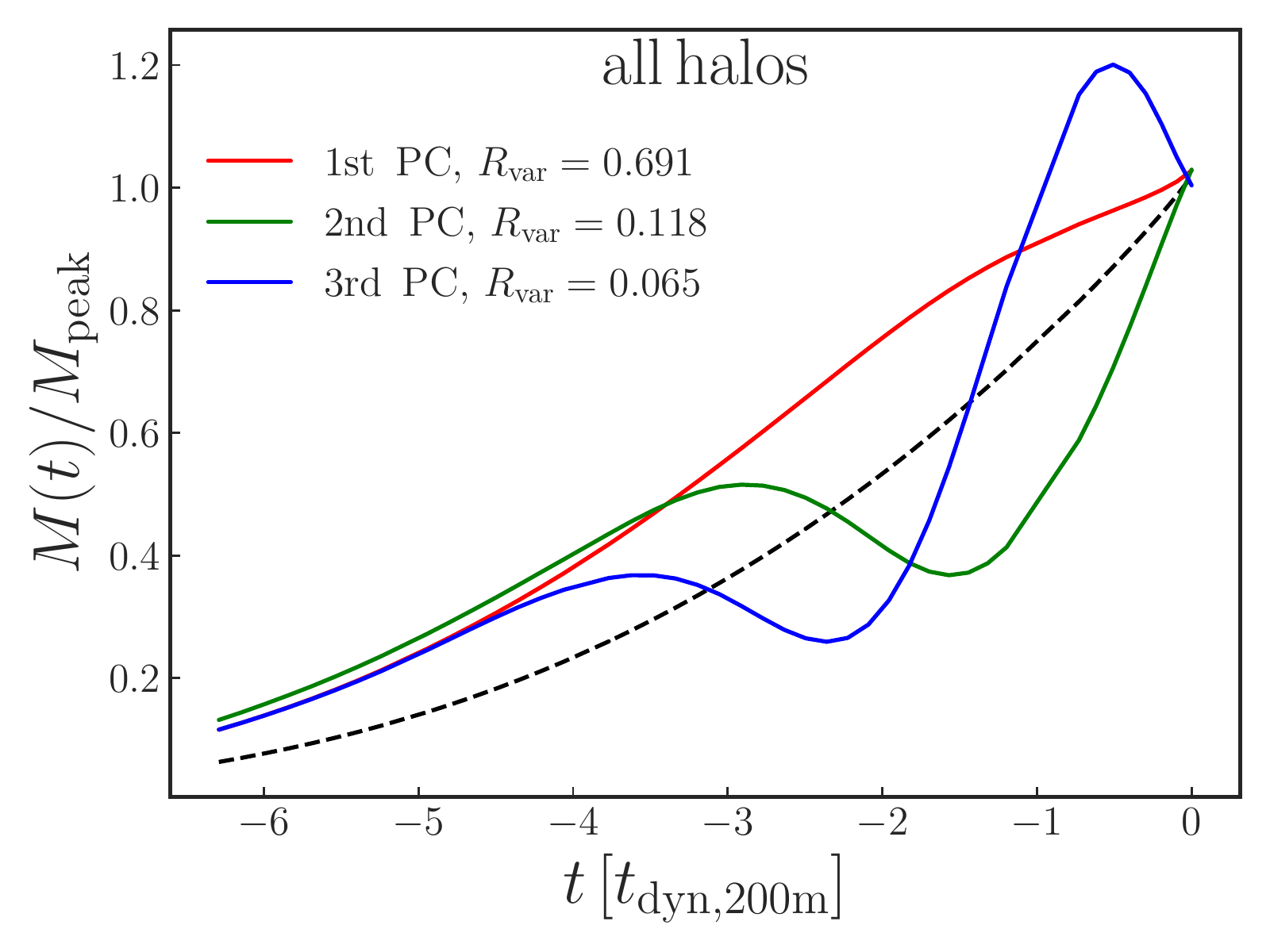}
	\caption{The first three principal components (red, green, blue) of the fractional mass history $M(t)/M_{\rm peak}$ (black dashed). $R_{\rm var}$ represents the explained variance ratio of each component.
	}
	\label{fig:Mz_PCA}
\end{figure}

In Section~\ref{sec:corr_rsp_mar:mah}, we explored the correlation between $\rsp$ and different epochs in the average mass accretion history. In this section, we explore whether we can explain the correlations based on a simplified version of individual halos' mass histories that are decomposed via a Principal Component Analysis (PCA). Since the mass history (first panel of Fig.~\ref{fig:M_history}) contain nearly identical information as the mass accretion history (third panel), we perform the PCA only on the mass history, $M(t)/M_{\rm peak}$. For $n$ data points (haloes) each having $m$ variables (epochs in their mass history), the PCA finds the vector in $m$-dimensional space along which the variance of the data points is maximised (the so-called `first principal direction'). The process is repeated for each subsequent dimension such that the new vector is orthogonal to all previous vectors. This procedure results in a vector space where the data can be exactly expressed as a series of coefficients times the component vectors, with the goal of reducing the dimensionality to the first few components (which ideally contain much of the `explained variance'). In other words, any halo's mass history can be accurately approximated as a linear combination of the first few PCA components. 

Fig.~\ref{fig:Mz_PCA} shows the first three PCA components of the mass history, including the explained variance ratios of about 69\%, 12\%, and 7\% (almost 90\% total). Unsurprisingly, adding more components adds little to the accuracy of the PCA. The first component (red) modulates the overall shape of the mass history, whereas a high second (green) component indicates strong recent mass accretion. In contrast, the third (blue) component does not clearly indicate an excess or lack of recent accretion.

\begin{figure*}
	\includegraphics[width=0.98\linewidth]{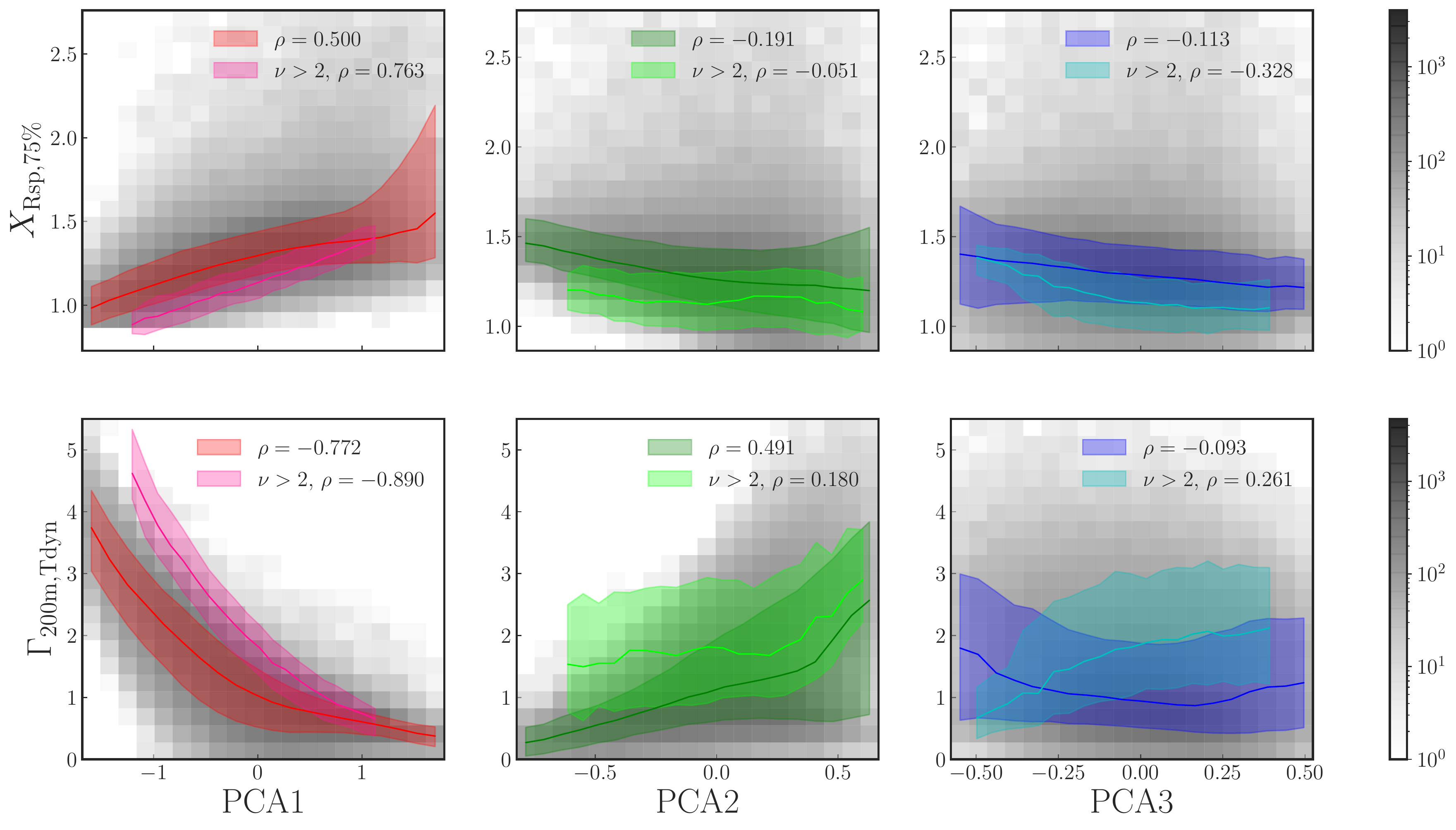}
	\caption{Relationship of the coefficients of the first three PCA components for each halo to their relative splashback radius (top) and mass accretion rate (bottom). The coloured lines and shaded areas show the median relationships and 68\% scatter. The first component carries most of the correlation to the splashback radius, which is not surprising since it is tightly correlated to the accretion rate. The second component exhibits weaker trends. The third component is barely correlated to the accretion rate (for the full halo sample), which makes sense given its shape within the past dynamical time (Fig.~\ref{fig:Mz_PCA}).
	}
	\label{fig:corr_PCA}
\end{figure*}

In Fig.~\ref{fig:corr_PCA} we investigate the relationship between the PCA coefficients of haloes and their relative splashback radii (top panels). We also explore how much of the correlation is mediated by the recent accretion rate by showing the correlation with $\Gamma$ in the bottom panels. We quantify the strength of the correlation with Pearson's coefficient (see legend). A high first coefficient is tightly correlated with a low accretion rate and thus with a high splashback radius (left panels of Fig.~\ref{fig:corr_PCA}). The correlation is particularly clear in the high-mass sample (pink line), presumably because large haloes dominate their environment and are thus less susceptible to other factors that might modulate $\rsp$. We find weaker and opposite trends in the second PCA component, where a high coefficient corresponds to a high recent accretion rate (middle panels). The third component is not predictive of recent accretion (Fig.~\ref{fig:Mz_PCA}) and also correlates with $\rsp$ rather weakly (right panels of Fig.~\ref{fig:corr_PCA}), although it is a stronger predictor of $\rsp$ in the high-mass sample. Overall, our results affirm once again that the majority of the effect on the splashback radius is mediated via the recent accretion over one dynamical time, and that the full mass history adds relatively little information.

\begin{figure}
	\includegraphics[width=0.98\linewidth]{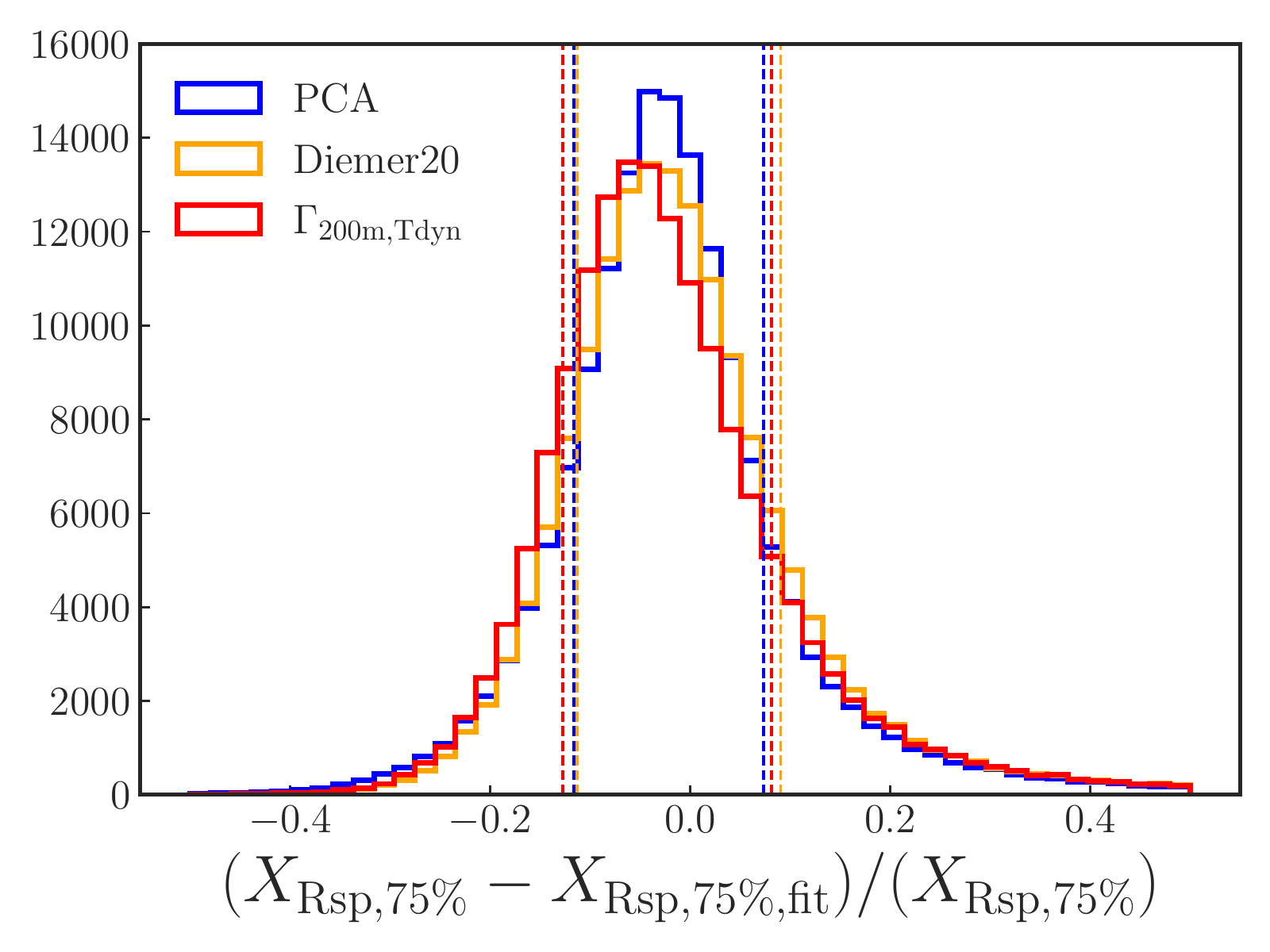}
	\caption{Fractional residuals from three different predictions for $R_{\rm sp,75\%}/R_{\rm 200m}$, namely, the median relation with the accretion rate (red), the fitting function of \citet[][yellow]{diemer_20_catalogs}, and the new polynomial fit based on the first three PCA components. The vertical dashed lines correspond to the 16\% and 84\% percentile values for each fit. The latter two predictions do not perform significantly better than the relationship with the accretion rate, but there is some additional information in the full shape of the mass history.
	}
	\label{fig:PCA_fit}
\end{figure}

We further assess the additional information content by constructing a simple fitting function that predicts the splashback radius from the PCA coefficients. In particular, we model $X_{\rm Rsp,75\%}$ with three second-order polynomials,
\begin{equation}
\label{eq:pca_fit}
    X_{\rm Rsp,75\%} = \left[\sum_{i=0}^2 (a_{x,i} x^i)\right]\left[\sum_{i=0}^2 (a_{y,i} y^i)\right]\left[\sum_{i=0}^2 (a_{z,i} z^i)\right] \,,
\end{equation}
where $x$,$y$ and $z$ are the coefficients of the first, the second and the third principal components, and $a_{x,i}$, $a_{y,i}$ and $a_{z,i}$ are free parameters. We determine the best-fit values of the 9 parameters by minimising $\chi^2$. For comparison, we construct an even simpler second-order polynomial `model' that is a function of only the accretion rate,
\begin{equation}
\label{eq:gamma_fit}
    X_{\rm Rsp,75\%} = \sum_{i=0}^2 (a_{\Gamma,i} \Gamma_{\rm 200m,Tdyn}^i).
\end{equation}
We quantify the success of these functions at predicting $\rsp$ in Fig.~\ref{fig:PCA_fit}, which shows the distribution of the fit residuals. While the PCA-based fit results in slightly lower average residuals than the purely accretion rate-based fit, the difference is marginal, especially given that eq.~\ref{eq:pca_fit} contains 9 free parameters and eq.~\ref{eq:gamma_fit} only 3. We also compare to the fitting function of \citet{diemer_20_catalogs} in yellow, which leads to slightly higher residuals than the PCA fit and a comparable performance eq.~\ref{eq:gamma_fit}. We note that the fitting also quantifies trends with $\nu$, $\Omega_{\rmm}(z)$, and percentile, and thus sacrifices some accuracy for a limited number of free parameters. 

In summary, the PCA provides an efficient way to parameterise mass histories. The first component captures $\sim70\%$ of the halo-to-halo fluctuations and is tightly correlated to the splashback radius ($\rho=0.763$ for the high-mass sample). This type of analysis could be applied to observational data, for example to constrain the accretion history of individual haloes based on their splashback radii (assuming those can be measured, e.g., from stellar halos; \citealt{deason_21}).

\bsp	
\label{lastpage}
\end{document}